\documentclass[lenghtcheck,twocolumn,subeqn,amsmath,floats,tabularx,unsortedaddress,superscriptaddress,nofootinbib,prl,amssymb,showpacs]{revtex4-1}
\usepackage{graphicx}%
\usepackage{dcolumn}%
\usepackage{bm}%
\usepackage{tabularx}  
\usepackage{multirow}
 
\usepackage{graphicx}
\usepackage{color}
\usepackage{amsmath}

\usepackage[utf8]{inputenc}

\begin{document}
\hyphenation{mo-des}
\title{Control of Recoil Losses in Nanomechanical SiN Membrane Resonators}

\author{A. Borrielli}
\affiliation{Institute of Materials for Electronics and Magnetism, Nanoscience-Trento-FBK Division, 38123 Povo (TN), Italy}
\affiliation{Istituto Nazionale di Fisica Nucleare, TIFPA, 38123 Povo (TN), Italy}

\author{L. Marconi}
\affiliation{CNR-INO, L.go Enrico Fermi 6, 50125 Firenze, Italy}
\affiliation{INFN, Sezione di Firenze, Via Sansone 1, 50019 Sesto Fiorentino (FI), Italy}

\author{F. Marin}
\affiliation{CNR-INO, L.go Enrico Fermi 6, 50125 Firenze, Italy}
\affiliation{INFN, Sezione di Firenze, Via Sansone 1, 50019 Sesto Fiorentino (FI), Italy}
\affiliation{Dipartimento di Fisica e Astronomia, Universit\`a di Firenze, Via Sansone 1, 50019 Sesto Fiorentino (FI), Italy}
\affiliation{LENS, Via Carrara 1, 50019 Sesto Fiorentino (FI), Italy}
\author{F. Marino}
\affiliation{CNR-INO, L.go Enrico Fermi 6, 50125 Firenze, Italy}
\affiliation{INFN, Sezione di Firenze, Via Sansone 1, 50019 Sesto Fiorentino (FI), Italy}
\author{B. Morana}
\affiliation{Institute of Materials for Electronics and Magnetism, Nanoscience-Trento-FBK Division, 38123 Povo (TN), Italy}
\affiliation{Delft University of Technology, Else Kooi Laboratory, 2628 Delft, The Netherlands}

\author{G. Pandraud}
\affiliation{Delft University of Technology, Else Kooi Laboratory, 2628 Delft, The Netherlands}
\author{A. Pontin}
\affiliation{INFN, Sezione di Firenze, Via Sansone 1, 50019 Sesto Fiorentino (FI), Italy}
\affiliation{Dipartimento di Fisica e Astronomia, Universit\`a di Firenze, Via Sansone 1, 50019 Sesto Fiorentino (FI), Italy}
\author{G.A.  Prodi}
\affiliation{Istituto Nazionale di Fisica Nucleare, TIFPA, 38123 Povo (TN), Italy}
\affiliation{Dipartimento di Fisica, Universit\`a di Trento, 38123 Povo (TN), Italy}
\author{P.M. Sarro}
\affiliation{Delft University of Technology, Else Kooi Laboratory, 2628 Delft, The Netherlands}
\author{E. Serra}
\affiliation{Istituto Nazionale di Fisica Nucleare, TIFPA, 38123 Povo (TN), Italy}
\affiliation{Delft University of Technology, Else Kooi Laboratory, 2628 Delft, The Netherlands}
\author{M. Bonaldi}
\email[Email:]{bonaldi@science.unitn.it}
\affiliation{Istituto Nazionale di Fisica Nucleare, TIFPA, 38123 Povo (TN), Italy}
\affiliation{Institute of Materials for Electronics and Magnetism, Nanoscience-Trento-FBK Division, 38123 Povo (TN), Italy}

\begin{abstract}
In the context of a recoil damping analysis, we have designed and produced a membrane resonator equipped with a specific on-chip structure working as a "loss shield" for a circular membrane. In this device the vibrations of the membrane, with a quality factor of $10^7$, reach the limit set by the intrinsic dissipation in silicon nitride, for all the modes and  regardless of the modal shape, also at low frequency. Guided by our theoretical model of the loss shield, we describe the design rationale of the device, which can be used as effective replacement of commercial membrane resonators in advanced optomechanical setups, also at cryogenic temperatures. 
\end{abstract}


\maketitle
 
Since the first demonstrations of use in optical cavity \cite{ThompsonNature2008,WilsonPRL2009}, membrane resonators have widely spread in optomechanical experiments, both as isolated mechanical oscillators and as components of hybrid systems. Their striking optical and mechanical properties allowed  the observation of quantum effects induced by optomechanical interaction in the behavior of nano-oscillators \cite{Purdy_Science_2013} and in the properties of radiation itself \cite{Purdy_PRX2013}. 

Currently, membrane based resonators represent a flexible tool for a wide range of scientific and technological goals: interfacing radiation at very different wavelengths \cite{AndrewsNphys2014,BagciNature2014}, implementing  hybrid mechanical–atomic systems \cite{TreutleinNNano2015}, fixing significant constraints on quantum gravity theories \cite{bawajNComms2015}, and studying multimode optomechanical systems in the quantum regime \cite{NielsenArxiv16}.
These developments motivate a strong commitment to improving the performance of membranes based oscillators. We address here the issue of mechanical losses in high stress silicon nitride (SiN) membranes, proposing a perspective which allows us to realize a "loss shield" for the membrane resonator. 

SiN membrane based devices have many mechanical resonances with frequencies starting from  $0.1\,$MHz, with intrinsic losses well described by a model \cite{VillanuevaPRL2014} where the elastic constant $K$ includes an imaginary part, $K=k(1+i \phi)$, with $\phi=1/Q$  the loss angle and $Q$  the quality factor. Though the intrinsic quality factor is in the range $10^6-10^8$, depending on dimensions and temperature, the loss through the supporting substrate can reduce this figure by several orders of magnitude. 
This phenomenon is more pronounced for the lower frequency resonances, which would be the most suitable for the experimental optomechanics as higher order resonances are surrounded by numerous neighboring resonances \cite{nearby}. 
An additional problem is the poor reproducibility of the loss contributed by the support, that depends on the mounting details \cite{PurdyNJP2014} and by the loss in the sample holder \cite{SchmidPRB2011}. It is known that these losses can sometimes be reduced by minimizing the contact of the chip frame with the sample mount, but this strategy is not optimal as it affects mechanical stability, position control and thermal anchoring of the membrane. 
 
The loss through the supporting substrate is usually evaluated from the energy transfer rate mediated by phonons tunneling from the membrane resonator into the substrate \cite{wilson-raePRB2008,Wilson-RaePRL2011}. However this theory cannot provide the guidelines for the design of more effective supporting systems, as it is based on some strong assumptions. In fact the substrate is described as an infinite half-plane, so its real modal structure is not considered. Moreover, the energy transfer is unidirectional, from the membrane to the substrate. Consequently, the loss calculated within this framework does not contain the loss angle of the mechanical resonances of the substrate, and therefore it is not expected any penalty by the use of lossy materials.  

In this paper we rely on a coupled oscillators model, in which the vibrations of the membrane are naturally combined with those of the support to give extended normal modes. Consequently the displacement field is distributed throughout the system and the loss at the membrane resonant frequency may be substantially degraded by the contribution of the support. This effect is called recoil damping and was first studied in suspension systems for interferometric detectors of gravitational waves \cite{SaulsonPRD1990}. In this context the effective quality factor $\overline{Q}_m$ of a membrane resonator supported by a wafer can be approximated as:
\begin{equation}
(\overline{Q}_m)^{-1}=Q_m^{-1}+Q_w^{-1}\frac{M_m}{M_w}\,\frac{\omega_w^2\omega_m^2}{(\omega_w^2-\omega_m^2)^2}
\label{eq:saulson}
\end{equation}
where $Q_m,\,M_m,\,\omega_m$ ($Q_w,\,M_w,\,\omega_w$) are the intrinsic quality factors, mass and resonant angular frequency of the membrane (supporting wafer).
According to this equation, a strong reduction of the quality factor is expected when a resonance of the support wafer approaches the resonance of the membrane, and the quality factor of the wafer oscillator $Q_w$ contributes to set the effective quality factor, in agreement with the experimental data reported in Refs. \cite{JockelAPL11} and  \cite{SchmidPRB2011}. 
If we consider that the recoil force due to the membrane is just the oscillating momentum $M_m \ddot{x}_{\mathrm{cm}}$, where $x_{\mathrm{cm}}(t)$ is the position of the center of mass of the membrane, we can easily understand another aspect of the behavior of membrane based resonator. In fact, it is common to observe a high quality factor for high frequency (i.e. high order) vibrations in commercial membranes. This is due to the increasingly large number of nodal lines, which averages out the contribution of different parts of the membrane in the evaluation of the center of mass, causing a strong reduction of the oscillating momentum and therefore of the recoil losses. We note that in modal shapes with an equal number of nodes and antinodes the center of mass is at rest, therefore these modes couple with the supporting wafer only through angular momentum, and the same reasoning can be repeated with moment of inertia in place of mass and recoil torque in place of recoil force.

\begin{figure}[!t]
\includegraphics[width=86mm]{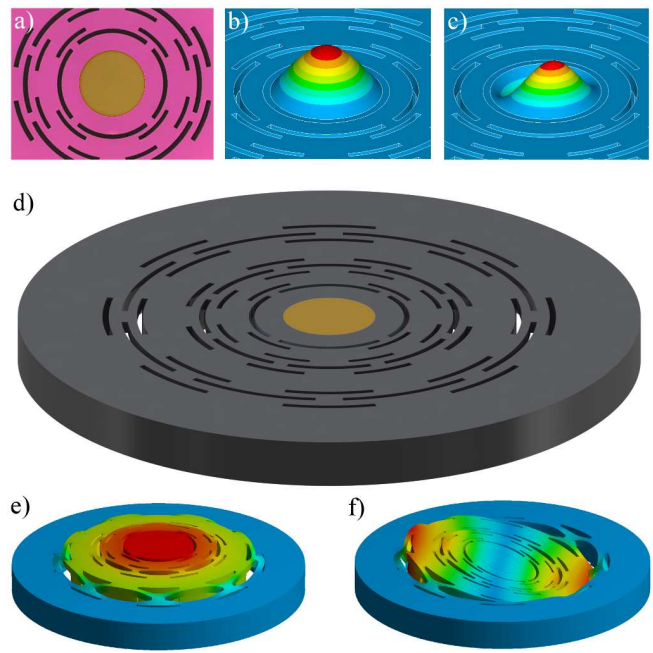}
\caption{(Color online) a) optical microscope picture of the circular membrane, with diameter $1.6\,$mm and thickness $100\,$nm. \mbox{b) - c)} First modal shapes involving the membrane, resonating at about $230\,$kHz and $366\,$kHz, respectively similar to the constrained membrane's normal mode $(n,k)=(0,1)$ and $(n,k)=(1,1)$. d) CAD image of the device. \mbox{e) - f)} Modal shapes of the lowest frequency resonances of the device, respectively at  32 kHz and 47 kHz, where it can be seen the planar displacement of the membrane frame.
\label{fig:device}}    
\end{figure}

Given that this approach reproduces the main features of the system, we have extended this theory to treat a multi-oscillator case, developing a scheme in which the contribution of the support to the effective quality factor of the membrane is greatly reduced. Following this design we have then produced a membrane resonator equipped with a specific on-chip structure working as a "loss shield" for the membrane. In this device also the low frequency vibrations of the membrane have a high quality factor and reach the limit set by the intrinsic dissipation.
Figure \ref{fig:device}a shows an optical microscope picture of the SiN membrane, with diameter $1.64\,$mm, thickness $100\,$nm and internal stress $\sim 0.8\,$GPa. This choice of thickness allows for a nearly optimal optical coupling at a wavelenght of $1064\,$nm \cite{wolf}. The membrane is supported by a silicon cylinder of diameter 2.4 mm and thickness $1\,$mm. This frame is supported in 4 points by a structure made of alternating flexural and torsional springs with thickness 250$\,\mu$m. This allows an oscillatory motion of the cylinder with minimal deformation during the displacement, in order to reduce the coupling of the internal resonances of the cylinder with the rest of the structure \cite{SerraAPL2012,BorrielliPRApplied2015}. As shown in Figure \ref{fig:device}d, this module cylinder-springs  is repeated 2 more times to obtain the desired degree of mechanical rigidity.
In Figures \ref{fig:device}e-f we show the planar displacement of the lowest frequency resonances of this structure, respectively at  $32\,$kHz and $47\,$kHz. The outer frame of the device, with thickness $1\,$mm, has a square shape $14\times 14\,$mm$^2$ and is clamped in a metallic holder with a central hole, which leaves uncovered the circular area of diameter  $10\,$mm shown in Figure \ref{fig:device}d.
The device is realized by through thickness fabrication of a silicon-on-insulator wafer with standard MEMS technology \cite{BorrielliPRApplied2015,SerraArxiv2016}. 

In a constrained circular membrane with a high intrinsic tensile stress  $\sigma_m$,  the theoretical resonance frequencies are given by the expression $f_{nk} = f_0\,\alpha_{nk}$ where $\alpha_{nk}$ is the $k$-th root of the Bessel polynomial of order $n$, and $f_{0} = \frac{1}{2 \pi R}\, \sqrt{\frac{\sigma_m}{\rho_m}}$, with $\rho_m$ the density and $R$ the radius of the membrane. The index $n$ may assume the values (0,1,\,..) and sets the number of nodal diameters of the normal mode, the index $k$ may assume the values (1,2,\,..) and sets the number of nodal circumferences. The Finite Elements shapes shown in Figure \ref{fig:device}b-c are nearly equivalent to the normal modes $(n,k)=(0,1)$ and  $(n,k)=(1,1)$ of the constrained membrane. 
We note that the effective modal mass of the $(0,k)$ modes for a centered, $\delta$–like readout decreases at higher values of $k$, because the absolute displacement of the modal shapes becomes more concentrated in the center  \cite{SerraArxiv2016}. In an optomechanical setup it is therefore possible to improve the coupling with the light by focusing the laser beam at the center of the membrane, implementing a nearly optimal readout. For comparison, in a square membrane the effective mass remains constant as the modal indexes change.

\begin{figure}[!t]
\includegraphics[width=86mm]{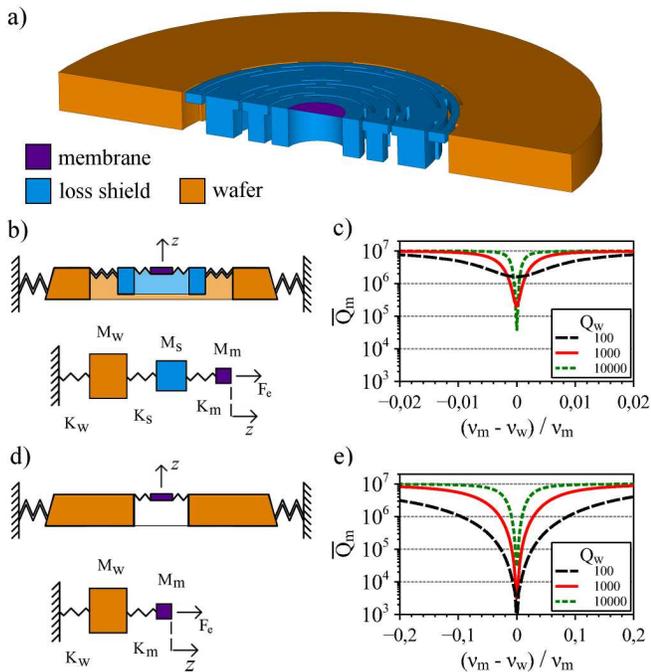}
\caption{(Color online) a) Section of the device where different colors denote the membrane resonator, the support wafer and the intermediate oscillator acting as loss shield. b) Schematic of the mechanical system and pointlike mass model. c) Plot of the main membrane mode effective quality factor $\overline{Q}_m$ when a wafer mode located at a frequency $\nu_w$ close to the membrane resonance $\nu_m$. d) Schematic of the mechanical system and pointlike mass model of a device without loss shield structure. e) Plot of the main membrane mode effective quality factor $\overline{Q}_m$ without loss shield, showing an "influence band" about 30 times larger. 
\label{fig:composita}}    
\end{figure}

To evaluate the effective quality factor of the membrane in the strong coupling model, we first identify three mechanical oscillators in the system: the  membrane resonator, the supporting wafer and an intermediate oscillator said loss shield, as shown in Figure \ref{fig:composita}a. Each oscillator is described by its frequency $\nu_i$ and mass $M_i$, where the subscript $i$ can be $(m, s, w)$ to identify respectively membrane, shield or wafer oscillators. We describe the loss in each part  by an imaginary component of the spring constant, $K_{i}=k_{i}(1+i/Q_{i})$, where $k_i=4 \pi^2\nu_i^2 M_i$ and $Q_{i}$ is the quality factor assigned to each oscillator. For a high-stress membrane of this size the intrinsic quality factor is $Q_m\simeq 10^7$, while the loss properties of the clamped area are determined by the losses induced by the sample holder, and it is common to observe $Q_w$ in the range $10^{2} - 10^{3}$. The quality factor $Q_s$ of the intermediate structures can vary from $10^3$ to some $10^4$, depending on the vibrational shape. 
In Figure \ref{fig:composita}b  we see the schematic of the device and the pointlike mass model used for the evaluation of the mechanical loss. The corresponding vibrational shapes of membrane and shield can be seen in Figures \ref{fig:device}b and \ref{fig:device}e respectively.
Given that in the coupled system it is no longer possible to distinguish the individual oscillators,  the membrane resonator is identified exactly as in an experiment, that is looking for the vibration of the system with frequency and effective mass closest to that expected. We stress that this method estimates exactly the quality factor measured in a dynamic experiment (response to an excitation or free decay) or in the thermal noise spectrum.
We define "coupled membrane" this observed system, while the membrane uncoupled oscillator corresponds to the normal mode of the constrained circular membrane.
Therefore we calculate the dynamic response of the system to a harmonic forcing $F_e$ applied to the membrane, and assess the quality factor $\overline{Q}_m$ from the linewidth \cite{Suppl}.

In the graph of Figure \ref{fig:composita}c we used the typical parameters of the $(0,1)$ membrane oscillator ($\nu_m=250\,$kHz,  $M_m=1.5\times 10^{-10}$kg, $Q_m=10^7$) and of the shield oscillator ($\nu_s=30\,$kHz, $M_s=1\times 10^{-5}\,$kg, $Q_s=10^3$). 
Instead we vary the frequency of the wafer oscillator, with mass $M_w=5\times 10^{-5}$kg, in the neighbourhood of $\nu_m$, where $\Delta \nu/\nu =(\nu_m-\nu_w)/\nu_m$ is the relative frequency shift. We assign three different values to the loss angle of the wafer  $1/Q_w$. It is evident that, until the wafer oscillator frequency is far enough  from the membrane frequency, the quality factor $\overline{Q}_m$ remains very close to the value assigned $Q_m$. The loss angle of the wafer oscillator becomes important when the two frequencies get closer.
In fact we see that with $Q_w=10^3$ we have $\overline{Q}_m/Q_m<0.9$ (corresponding to a 10\% decrease of the quality factor) when $\Delta \nu/\nu  <0.01$, but with $Q_w=10^2$ we have the same decrease when $\Delta \nu <.035$. The effectiveness of this design is evident from the comparison with the standard situation, without loss shield and the membrane directly supported by the wafer (Fig. \ref{fig:composita}d).  As shown in Figure \ref{fig:composita}e, with  $Q_w=10^3$, we have $\overline{Q}_m/Q_m<0.9$ within a quite large influence band $\Delta \nu/\nu <0.3$, that enlarges to $\Delta \nu/\nu  <0.6$ if $Q_w=10^2$. This figure makes clear that a wafer oscillator with very low  $Q_w$ can increase the loss of the membrane oscillator even if it is relatively distant in frequency. Since the quality factor of the wafer's vibrations is determined also by the sample holder, this recoil losses model explains the great influence of the assembly and the reproducibility issues. 
  
This reasoning can be extended to all of the modal shapes of the coupled membrane oscillator, as the device naturally profits by the various modes of the intermediate elastic structure: for instance, the resonance shown in Fig. \ref{fig:device}c couples with the support through recoil torque and is shielded by the mode of Fig. \ref{fig:device}f. This ensures an efficient shielding for all of the resonances of the coupled membrane, starting with the lowest frequency. However the occurrence of wafer and structure modes with frequency coinciding to the resonances of the coupled membrane should be avoided, given that in this case the shield does not completely eliminate the coupling loss. In any case the probability of spoiling a substantial number of coupled membrane resonances is much reduced by the use of the filter, thanks to the reduction by a factor of 30 of the influence band.
 
\begin{figure}[!t]
\includegraphics[width=86mm]{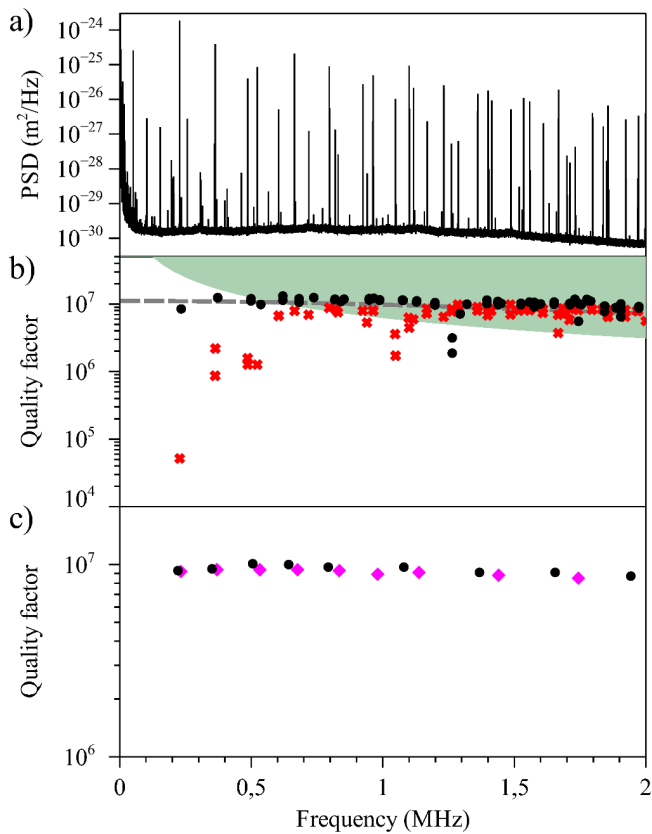}
\caption{(Color online) a) Power Spectrum Density of the displacement noise of the membrane. Some electrical and seismic lines can be seen especially below 1 MHz. The amplitude of coupled membrane resonances is compatible with thermal force noise driving.
b)  Measured quality factor of each resonance. (closed circles) main device; (crosses) test device with reduced loss shield. The dashed line shows the quality factor envelope due to intrinsic loss for a square membrane with side $1.52\,$mm and fundamental frequency $230\,$kHz, evaluated \cite{YuPRL2012} with a loss angle $\phi=1.25\times 10^{-4}$. The data points within the shaded area fulfill the minimum requirement for quantum optomechanics at room temperature, $\overline{Q}_m>6\times 10^{12}/\nu_m $ \cite{optomechanicsRMP2014}. c) Quality factor at $4.3\,$K for two samples, with slightly different intrinsic stress. At this temperature the requirement for quantum optomechanics is surpassed by more than one order of magnitude.
\label{fig:misure}}    
\end{figure}

The coupled membrane resonances can be seen in the thermal Power Spectrum Density of the membrane displacement  (Fig. \ref{fig:misure}a), acquired using a Michelson interferometer  with the sample kept in a vacuum chamber. 
The frequency of all resonances are in good agreement with the modal frequencies of the constrained membrane, if we set $\sigma_m/\rho_m=2.4\times 10^5\,$Nm/kg. 
To measure the quality factor we drive the system by a piezoelectric actuator mounted on the sample holder and measure the free decay time of each coupled membrane resonance. In Figure  \ref{fig:misure}b we plot, for the main device shown in Figure \ref{fig:device}, the quality factor of all resonances up to a frequency of  $2\,$MHz. All vibrations with frequency higher than $0.5\,$MHz fulfill the minimum requirement for quantum optomechanics at room temperature, $\overline{Q}_m>6\times 10^{12}/\nu_m $ \cite{optomechanicsRMP2014}. 
We show also the quality factor of a test device with a membrane of the same size but featuring a shield oscillator resonating at about $90\,$kHz (about three times the value chosen for the main device shown in Figure \ref{fig:device}). As explained in the Supplemental Material \cite{Suppl}, this shield is much less effective in preserving the intrinsic quality factor of the membrane. 
In both cases the results confirm our recoil losses analysis. For the main device, the quality factor measured at low frequency is compatible with the value calculated  for a square membrane with the same fundamental frequency \cite{YuPRL2012,VillanuevaPRL2014}. In Figure \ref{fig:misure}c we show the quality factor of some coupled membrane resonances at $4.3\,$K, corresponding to (0,k) and some (1,k) modes of the constrained membrane.
These results confirm the predictive value of our model and its practical effectiveness. In comparison with unshielded membranes \cite{YuPRL2012,chakramPRL2014}, our device reaches the limit set by the intrinsic dissipation starting from the low frequency resonances and regardless of the modal shape, independently from the experimental setup and with a clamping system suitable for the cryogenic use. 

Within the recoil losses framework we can also estimate the quality factor of membranes with phononic shields, where the wafer becomes a periodic structure with bandgaps \cite{YuAPL2014,TsaturyanOE2014}. In this case membrane vibrations with frequency within the bandgap have demonstrated high quality factors independently of the clamping system. In our context, with a bandgap of about $0.3\,$MHz, the frequency shift between a membrane resonance at $3\,$MHz and structure modes is $\Delta\nu/\nu\simeq 0.05$. As a rough estimate, from Figure \ref{fig:composita}e we can expect a reduction of about 50\% of the quality factor in the case $Q_w=1000$, which it is a good result but still not optimal. However, better results can be obtained by widening the bandgap or by cooling the sample at cryogenic temperatures, where the influence bands shrink thanks to a general improvement in the wafer's quality factor. Unfortunately the extension of this technique to the $100\,$kHz range would require the use of very large isolation structures. In fact a bandgap centered at a frequency $f=100\,$kHz can be obtained with a $10\times 10$ array of unit cells with characteristic length $v/(2f)\simeq 10\,$mm, where $v$ is the sound velocity in silicon. 

We also mention recent devices where small membranes (about $0.1\times 0.1\,$mm$^2$) are supported by thin tethers \cite{SankeyPRX2016,NortePRL2016} that act as decoupling elements. These resonators feature a mass in the ng range, obtained at the expenses of a reduced thermal management capability, thus targeting different experimental setup in respect to large membranes. We note that in these devices the quality factor at room temperature can reach a value as high as $10^8$, thanks to the reduction of the contact area and the use of reduced thickness of the membranes down to $20\,$nm. This finding suggests that the use of thin membranes could improve the performances of our devices through the reduction of the intrinsic loss \cite{VillanuevaPRL2014}, although in this case the use of a nanostructured pattern \cite{NortePRL2016}  may become necessary to increase the reflectivity and restore the overall optomechanical coupling rate \cite{SankeyPRX2016}.

All thing considered, our recoil losses analysis allows to build robust devices with high quality factor, preserving the thermal and geometrical characteristics of typical membrane based resonators. The optical properties of these membranes are compatible with their use as optomechanical oscillators, both in Michelson interferometers and in cavity setups \cite{SerraArxiv2016}. 
Since their quality factor is high in the whole frequency range, they can be used with optimal efficiency, both in single-mode applications (such as optical cooling \cite{PetersonPRL2016} and force sensing \cite{SankeyPRX2016}) and in multi-mode applications (such as hybridization \cite{ShkarinPRL2014} and two-mode squeezing \cite{PatilPRL2015}). For these reasons we imagine that this class of devices may spread as effective replacement of standard membrane resonators in advanced optomechanical setups, also at cryogenic temperatures.

This work has been supported by MIUR ("PRIN 2010-2011" and "QUANTOM") and by INFN ("HUMOR" project). A.B. acknowledges support from the MIUR under the "FIRB-Futuro in ricerca 2013" funding program, project code RBFR13QUVI.


\begin{thebibliography}{99}
\bibitem{nearby}Nearby modes can interfere with the measurement by increasing the displacement background noise or by coupling with the main mode through the intracavity field.
\end{thebibliography}
\end{document}

\title{Supplemental material to:\\"Control of Recoil Losses in Nanomechanical SiN Membrane Resonators"}

\maketitle

%\clearpage

%\section{Design and assessment of the devices}
\begin{center}
\textbf{\large{Section I. Coupled modes model}}
\end{center}

\begin{figure}[b!]
\includegraphics[width=130mm]{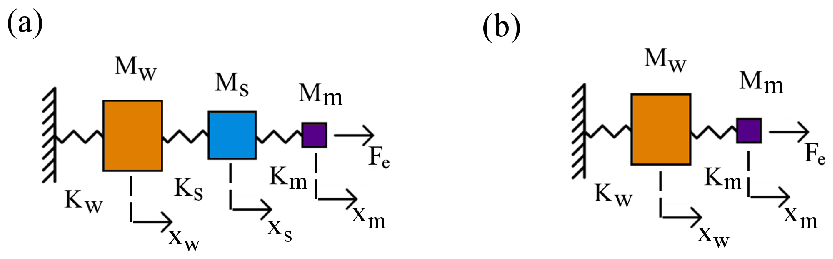}
\caption{Schematic models used to calculate the effect of wafer vibrations on the quality factor of the {coupled membrane}. The $(w,s,m)$ oscillators represents the wafer, the loss shield and the membrane, respectively. The imaginary part in the spring constant $K_i=k_i(1+i/Q_i)$, with $i=w,\,s,\,m$, describes the structural loss in the uncoupled oscillators.  
 \label{fig:toymodel}}
\end{figure}

In the frequency domain (here and in the following we indicate the Fourier transform with a tilde) the motion of the masses in the model with the loss shield oscillator, sketched in Figure \ref{fig:toymodel}a, is described by the equation:
\begin{equation}
\label{eq:3portimpedance}
 \left[Z^{(3)}_{ij}\right] \left( \begin{array}{c}\widetilde{x}_m\\\widetilde{x}_s\\\widetilde{x}_w\end{array} \right)
=
\left( \begin{array}{c}\widetilde{F}_e\\0\\0\end{array} \right)
\end{equation}
where $\widetilde{x}_i$, with $i=m,s,w$, are the displacements from the equilibrium position and:
{\small\begin{equation}
\label{eq:Z3}
\left[Z^{(3)}_{ij}\right]=\left[ \begin{array}{ccc}k_m\left(1+\frac{i}{Q_m}\right)-M_m\omega^2  & -k_m\left(1+\frac{i}{Q_m}\right) &0 \\-k_m\left(1+\frac{i}{Q_m}\right) & k_s\left(1+\frac{i}{Q_s}\right) +k_m\left(1+\frac{i}{Q_m}\right) -M_s \omega^2 &-k_s\left(1+\frac{i}{Q_s}\right)\\0 & -k_s\left(1+\frac{i}{Q_s}\right) & k_s\left(1+\frac{i}{Q_s}\right)+ k_w\left(1+\frac{i}{Q_w}\right)-M_w \omega^2 \end{array} \right]
\end{equation}
}
The formal solution of Eq. \eqref{eq:3portimpedance} is:
\begin{equation}
\label{eq:3portimpedance2}
 \left( \begin{array}{c}\widetilde{x}_m\\\widetilde{x}_s\\\widetilde{x}_w\end{array} \right)
=\left[Z^{(3)}_{ij}\right] ^{-1}
\left( \begin{array}{c}\widetilde{F}_e\\0\\0\end{array} \right)
\end{equation}

In a similar way, the mechanical response of the model with two oscillators (Figure \ref{fig:toymodel}b) is:
\begin{equation}
\label{eq:2portimpedance}
 \left[Z^{(2)}_{ij}\right] \left( \begin{array}{c}\widetilde{x}_m\\\widetilde{x}_w\end{array} \right)
=
\left( \begin{array}{c}\widetilde{F}_e\\0\end{array} \right)
\end{equation}
with:
\begin{equation}
\label{eq:twoportimpedance}
\left[Z^{(2)}_{ij}\right]=
\left[ \begin{array}{cc}-M_m\omega^2+k_m\left(1+\frac{i}{Q_m}\right)  & -k_m\left(1+\frac{i}{Q_m}\right) \\-k_m\left(1+\frac{i}{Q_m}\right) & -M_w \omega^2 +k_w\left(1+\frac{i}{Q_w}\right)+k_m\left(1+\frac{i}{Q_m}\right) \end{array} \right]
\end{equation}
and formal solution:
\begin{equation}
\label{eq:2portimpedance2}
  \left( \begin{array}{c}\widetilde{x}_m\\\widetilde{x}_w\end{array} \right)
=\left[Z^{(2)}_{ij}\right]^{-1}
\left( \begin{array}{c}\widetilde{F}_e\\0\end{array} \right)
\end{equation}

These equations could be used to calculate the resonance frequency and the system's normal modes, which, however, are complex given that the matrix that describes the elastic constants is not Hermitian. On the other hand it is not necessary to know the specific combination of displacements $\sum_i a_i x_i$ (with $a_i$ complex constants) that constitutes the normal modes of the compound system, and even the generalized forces (i.e. the combination of forces to be applied to individual masses to drive only one of these modes). In fact on a practical level we only need the response of the membrane  $M_m$ to an external force applied  $F_e$. This information makes it possible to predict the thermal noise and the response to the detection back-reaction, since the readout measures only the displacement  $x_m$.

From Equations \eqref{eq:3portimpedance2} and \eqref{eq:2portimpedance2} we easily obtain the transfer functions of the system relative to the applied force $F_e$:
\begin{eqnarray}
\label{eq:soluzioneTF}
\widetilde{x}_m=T_2 \widetilde{F}_e\\
\widetilde{x}_m=T_3 \widetilde{F}_e
\end{eqnarray}
If the resonant frequencies of the uncoupled oscillators ($\omega_i=\sqrt{k_i/M_i}$) are well separated and the loss terms are reasonably small (say $Q_i\geq 100$), it will be easy to identify three resonance peaks close to the resonances $\omega_i$.
Given that in the coupled system it is no longer possible to identify the individual oscillators,  the vibrations of the membrane are identified exactly as in an experiment, that is looking for the resonant peak of the system with frequency and effective mass closest to that expected. We stress that this method estimates exactly the quality factor measured in a dynamic experiment (response to an excitation or free decay) or in the thermal noise spectrum. We define "coupled membrane" this observed system, that must non be confused with the membrane uncoupled oscillator. In fact each uncoupled oscillators has a well defined quality factor depending on the imaginary part assigned to its spring constant $K_i$. For the membrane oscillator, the uncoupled oscillator represents a normal mode of the constrained circular membrane. 

%should not be used in our context, as it refers to a normal mode where only the membrane is oscillating. As we have noted above, this kind of oscillation is not a solution of the system's equations. Moreover this picture can be misleading, as it is difficult to understand why the quality factor of the membrane should be affected by the supporting frame if the displacement field is confined in the membrane itself. 

%\clearpage

%\newpage
\begin{center}
\textbf{\large{Section II. Approximate solution of the model with two coupled oscillators}}
\end{center}

For the two-oscillators system of Figure \ref{fig:toymodel}b, it is straightforward to show that:
\begin{eqnarray}
\label{eq:T2sol}
&&T_2=\\
&&-\frac{-M_w \omega^2 + M_w\omega_w^2(1+\frac{i}{Q_w})+ M_m \omega_m^2(1+\frac{i}{Q_m})}{M_m^2\omega_m^4(1+\frac{i}{Q_m})^2-\left( -M_m \omega^2 +M_m \omega_m^2(1+\frac{i}{Q_m})\right)\left( -M_w\omega^2  + M_w\omega_w^2(1+\frac{i}{Q_w})+ M_m \omega_m^2(1+\frac{i}{Q_m})   \right)}\nonumber
\end{eqnarray}
where $\omega_w=\sqrt{k_w/M_w}$ and $\omega_m=\sqrt{k_m/M_m}$ are the angular resonant frequencies of the uncoupled wafer and membrane oscillators, respectively.

From the resonance linewidth it is possible to calculate an approximate relationship \cite{SaulsonPRD1990} for the effective quality factor  $\overline{Q}_m$ of the {coupled membrane}:
\begin{equation}
\overline{Q}_m\,^{-1}=Q_m^{-1}+Q_w^{-1}\frac{M_m}{M_w}\,\frac{\omega_w^2\omega_m^2}{(\omega_w^2-\omega_m^2)^2}
\label{eq:saulson}
\end{equation}
This equation, proposed for evaluating the recoil losses in a chain of two pendulums like that used in interferometers for the detection of gravitational waves, describes the reduction of the quality factor that is observed when a resonance of the supporting wafer approaches the resonance of the membrane. Because of the approximations made in the derivation, the formula is not valid in the case $\omega_w \simeq \omega_m$. In fact in Eq. \eqref{eq:saulson}, when the two oscillators are in resonance the effective quality factor vanishes independently of other system parameters, such as the mass ratio $M_m/M_w$ and the quality factor of the wafer oscillator $Q_w$.

Even if the numerical approach is the only one able to give a response over the entire frequency band, the Equation \eqref{eq:saulson} can be complemented with an  approximation which is only valid in the case $\omega_w \simeq \omega_m$:
\begin{eqnarray}
&&T_2=\\
&&-\frac{1}{Q_w M_w \omega_w^2\left( \frac{M_m^2}{M_w^2} -2 \frac{M_m}{M_w} \left(\frac{\omega^2}{\omega_w^2}-1\right)\left(\frac{\omega^2}{\omega_m^2}-1\right)
+ \left(\frac{1}{Q_w^2}+\left(\frac{\omega^2}{\omega_w^2}-1\right)^2 \right) \left(\frac{1}{Q_m^2}+\left(\frac{\omega^2}{\omega_m^2}-1\right)^2 \right)  \right)}\nonumber
\end{eqnarray}
If $Q_m>>1$ and $M_w/M_m >> Q_w^2$ this equation can be used to evaluate the coupled membrane's effective quality factor as: 
\begin{equation}
\overline{Q}_m\,^{-1}=Q_m^{-1}+ Q_w^2\frac{M_m^2}{M_w^2}
\label{eq:approxresonance}
\end{equation}

This relationship defines the interplay between the ratio of the masses $M_m/M_w$ and the quality factor of the wafer oscillator. In particular, we see that if the mass of the membrane is negligible, its quality factor remains practically unchanged. Also it is seen that in this regime a low value of $Q_w$  can improve the effective quality factor $\overline{Q}_m$. This behavior is confirmed by the numerical evaluation of the effective quality factor, valid across the whole range of frequencies and for all values of $M_i$ and $Q_i$.

%\clearpage
%\newpage
\begin{center}
\textbf{\large{Section III. Numerical solution}}
\end{center}

We have shown that in the two-oscillator system it is not available a general relation, valid for any value of the uncoupled frequencies $\omega_i$, for the evaluation of the quality factor $\overline{Q}_m$ of the coupled membrane.

We therefore obtain a numerical estimate of the quality factor from the imaginary part of the transfer function. This is formally equivalent to study the thermal noise PSD (power spectrum density) of the resonator, that  is  predicted  by  the  fluctuation-dissipation theorem \cite{FTD} to be: 
\begin{equation}
S(\omega)= -\frac{4 k_B T}{\omega}\Im{[T(\omega)]}
\label{eq:FTD}
\end{equation}
This spectrum has peaks near the uncoupled frequencies of the various oscillators that make up the system. The coupled membrane resonator is identified by looking for the resonance of the system with frequency and effective mass closest to that expected. As said above,  this method estimates exactly the quality factor measured in a dynamic experiment (response to an excitation or free decay) or in the thermal noise spectrum. Therefore we assess the quality factor $\overline{Q}_m$ from the line width, on the basis of a comparison with the expected thermal noise in an uncoupled oscillator  with mass $\overline{M}_m$, angular frequency $\overline{\omega}_m$ and loss angle $1/\overline{Q}_m$. In this case the response to the external drive $F_e$ is: 
\begin{equation}
\widetilde{x}_m=T_1 \widetilde{F}_e
\end{equation}
where
\begin{equation}
T_1= \frac{1}{\overline{M}_m}\frac{1}{(\omega^2-\overline{\omega}_m^2)+i  \overline{\omega}_m^2/ \overline{Q}_m^2}
\label{eq:Lorentz}
\end{equation}
and the imaginary part of $T1$ is a Lorentzian function:
\begin{equation}
\Im{[T_1(\omega)]}= -\frac{1}{ \overline{M}_m }{\,\, }\frac{\overline{\omega}_m^2/\overline{Q}_m}{(\omega^2-\overline{\omega}_m^2)^2+{\overline{\omega}_m^4}/{\overline{Q}_m^2}}
\label{eq:imagLorentz}
\end{equation}
We point out that, in our numerical evaluation, the analysis of the imaginary part is numerically more efficient than the study of the transfer function module. In fact, in Eq. \eqref{eq:imagLorentz} both the width and the amplitude of the response are dependent on the quality factor.

\begin{figure}[h]
\includegraphics[width=100mm]{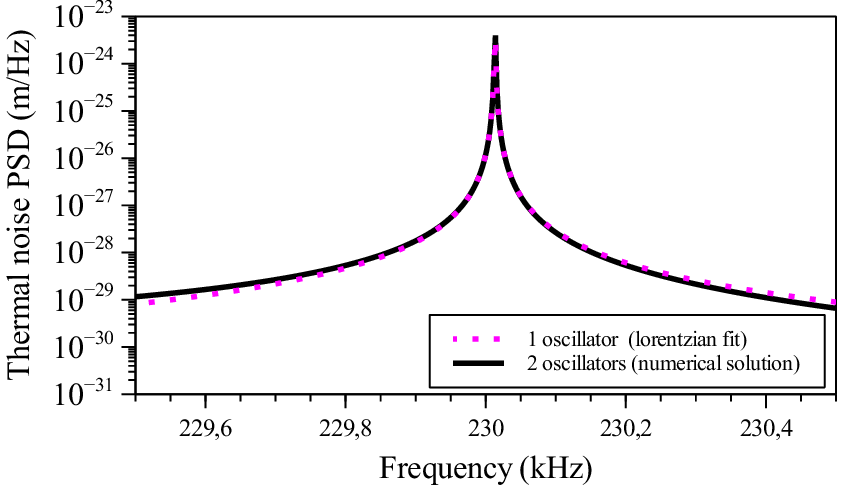}
\caption{Lorentzian fit of the numerical solution of thermal noise PSD for the two oscillator model.
 \label{fig:fit2modi}}
\end{figure}

Essentially the assessment of the coupled membrane consists in the fitting of the PSD predicted by models of Equations \eqref{eq:soluzioneTF}, with the Lorentzian function  \eqref{eq:imagLorentz}, to obtain the effective values of the angular frequency $\overline{\omega}_m$ and of the quality factor $\overline{Q}_m$. In Figure \ref{fig:fit2modi}  we see an example of Lorentzian fit. In this procedure we always make the assumption that the corrections to the membrane mass are negligible ($\overline{M_m}\simeq M_m$), given that $M_m << M_w$. Typically also the frequency $\overline{\omega}_m$  turns out to be the same of the uncoupled oscillator, while the quality factor $\overline{Q}_m$ can become smaller in magnitude than the uncoupled $Q_m$  when the frequency of the wafer oscillator approaches $\omega_m$.

\begin{figure}[h]
\includegraphics[width=140mm]{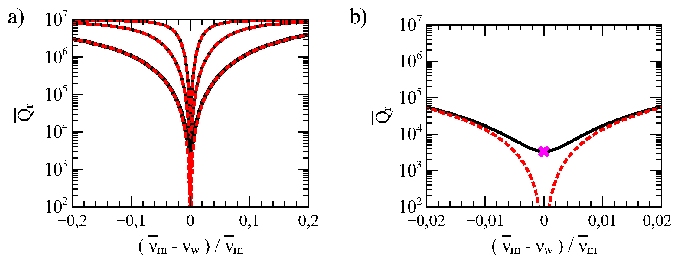}
\caption{Effective quality factor of the membrane oscillator $\overline{Q}_m$ as a function of the the relative frequency shift of $\nu_w$ from the effective frequency  $\overline{\nu}_m$.  a) Comparison between the numerical solution (solid line) and the approximated estimate of Equation \eqref{eq:saulson} (dotted red line) for three values of the quality factor  $Q_w=10^2,\,10^3,\,10^4$, from the upper curve to the lower curve. As expected the agreement is very good out of resonance $(\overline{\nu_m}-\nu_w)/\overline{\nu_m} >0.01$. b) Details of the behavior in the resonant case $(\overline{\nu_m}-\nu_w)/\overline{\nu_m} <0.01$ and $Q_w=10^2$. The numerical solution (solid line) is consistent with the estimate Eq. \eqref{eq:approxresonance} (magenta cross), valid at resonance when $Q_w<<\sqrt{M_w/M_m}\simeq 500$. We also plot  the approximated solution given by Equation \eqref{eq:saulson}  (dotted red line).
\label{fig:saulson}}
\end{figure}

To validate our numerical evaluation, we compare in Figure \ref{fig:saulson}, for the two oscillator model Fig. \ref{fig:toymodel}b, the numerical procedure and the approximate relationship Eq. \eqref{eq:saulson}. The membrane oscillator has mass $M_m=1.5\times 10^{-10}\,$kg and frequency $\omega_m/(2 \pi)=\nu_m=230\,$ kHz, with uncoupled quality factor $Q_m=10^7$. The wafer oscillator has mass $M_w=5\times 10^{-5}\,$kg and its frequency $\nu_w=\omega_w/(2\pi)$ is varied in the neighborhood of $\nu_m$. 
In Figura \ref{fig:saulson}a we show the quality factor of the coupled membrane  for three values of the quality factor  $Q_w$, from $10^2$ to $10^4$.
Here $(\overline{\nu_m}-\nu_w)/\overline{\nu_m}$ is the relative frequency shift of $\nu_w$ from the effective frequency $\overline{\nu}_m$.
Both approaches give the same result if the two resonances are separated by more than 1\%, in agreement with the assumptions underlying the derivation of Eq. \eqref{eq:saulson}. As shown in Figure \ref{fig:saulson}b, the numerical evaluation is fully consistent with our approximation of Eq. \eqref{eq:approxresonance} if $Q_w<<\sqrt{M_w/M_m}\simeq 500$.

%\clearpage
\newpage
\begin{center}
\textbf{\large{Section IV. Three coupled oscillators}}
\end{center}

The system with three oscillators is even more complex and  analytical estimates of  $\overline{Q}_m$ can be obtained only in very special cases, such as identical frequencies $\omega_i=\omega_0$ or identical losses $Q_i=Q_0$, that are not useful for our application.   We therefore rely on the numerical procedure, that reproduces very well the approximate theoretical estimates in the two oscillators case.

\begin{figure}[h!]
\includegraphics[width=162mm]{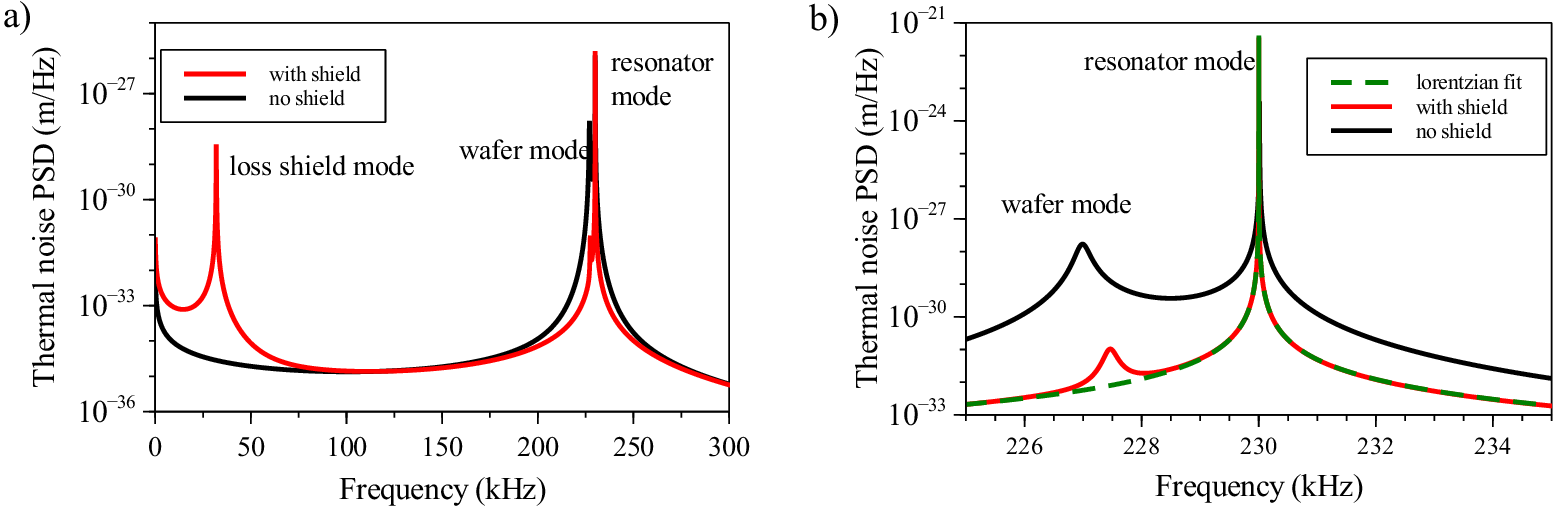}
\caption{PSD of the thermal noise predicted by the model. The uncoupled membrane and wafer resonances are respectively at $\nu_m=230\,$kHz and $\nu_w=227\,$kHz. The shield oscillator resonance is at $30\,$kHz.
 \label{fig:modelcomparison}}
\end{figure}

We show in Figure  \ref{fig:modelcomparison} the thermal noise PSD for a coupled membrane in the three- and two-oscillator model. 
The membrane oscillator has mass $M_m=1.5\times 10^{-10}\,$kg and frequency $\omega_m/(2 \pi)=\nu_m=230\,$ kHz, with uncoupled quality factor $Q_m=10^7$. The wafer oscillator has mass $M_w=5\times 10^{-5}\,$kg and frequency $\nu_w=227\,$kHz. The loss shield oscillator has frequency $\nu_s=30\,$kHz, mass $M_s=1\times 10^{-5}\,$kg and quality factor $Q_s=10^3$. 
In Figure \ref{fig:modelcomparison}a we show the full frequency range, where we can see the peak at $30\,$kHz due to the thermal motion of the shield oscillator.
In Figure \ref{fig:modelcomparison}b we show the detail of the resonance, where the contributions from the membrane and the wafer oscillators can be clearly seen.
The frequency of the coupled membrane is about the same with both models,  while the coupled wafer resonance shifts to $227.5\,$kHz in the three-oscillators case, due to the coupling with the shield oscillator.   
The use of the loss shield leads to a substantial narrowing of the coupled membrane linewidth and to a corrispondent improvement in the thermal noise level. The quality factor rises from $\overline{Q}_m\simeq 2\times 10^5$ to $\overline{Q}_m\simeq 8\times 10^6$ (obtained by the Lorentzian fit shown in Figure \ref{fig:modelcomparison}b).

%\clearpage
\begin{center}
\textbf{\large{Section V. Shielding effect and oscillator's parameters}}
\end{center}

The shielding effect of the three oscillator configuration does not depend critically on the parameters of the oscillators. This is useful because the different oscillators that we have described by pointlike masses and ideal springs are actually distributed in the structure and it would be difficult to obtain precise values.

\begin{figure}[h!]
\includegraphics[width=86mm]{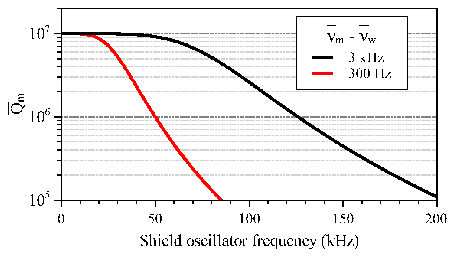}
\caption{Effective quality factor of the membrane as a function of the frequency  $\nu_s$ of the shield oscillator. The curve is evaluated for two values of the difference between the effective frequencies of the membrane oscillator $\overline{\nu}_m$ and of the wafer oscillator $\overline{\nu}_w$. We observe that $\overline{\nu}_m\simeq \nu_m=230\,$kHz in the full range studied, while the uncoupled frequency of the wafer $\nu_w$ must be varied to keep constant the coupled value $\overline{\nu}_w$ when the parameters of the shield oscillator are varied. This occurs because the effective masses of the wafer and shield oscillators are quite similar and the variation of the parameters can easily cause a shift of a few hundred Hz. 
 \label{fig:variazioneFs}}
\end{figure}

In particular, as regards the shield oscillator parameters, the effectiveness of shielding remains unchanged even by reducing or increasing the quality factor $Q_s$ by a factor of 10. Its frequency $\nu_s$ must instead remain well below the frequency of the membrane oscillator $\nu_m$. We show in Figure \ref{fig:variazioneFs} the efficiency of the shielding as a function of the frequency $\nu_s$.
For example, the effective quality factor $\overline{Q}_m$ of the coupled  membrane at $\overline{\nu}_m=230\,$kHz remains virtually undisturbed by the presence of the wafer oscillator at $\overline{\nu}_w=227\,$kHz, until the frequency of the shield oscillator remains below $40\,$kHz.

\clearpage

%\begin{center}
%\textbf{\large{Section V. Fabrication of the devices}}
%\end{center}
%\clearpage